\documentclass[aps,prl,twocolumn,amsmath,amssymb,showpacs,superscriptaddress]{revtex4-1}
\usepackage{graphicx}
\pdfoutput=1

\begin{document}
\title{Crossed Andreev Reflection in Quantum Wires with Strong Spin-Orbit Interaction}

\author{Koji Sato}
\affiliation{Department of Physics and Astronomy, University of California, Los Angeles, California 90095, USA}
\author{Daniel Loss}
\affiliation{Department of Physics, University of Basel, Klingelbergstrasse 82, CH-4056 Basel, Switzerland}
\author{Yaroslav Tserkovnyak}
\affiliation{Department of Physics and Astronomy, University of California, Los Angeles, California 90095, USA}

\begin{abstract}
\end{abstract}

\begin{abstract}
We theoretically study tunneling of Cooper pairs from an $s$-wave superconductor into two semiconductor quantum wires with strong spin-orbit interaction under magnetic field, which approximate helical Luttinger liquids. The entanglement of electrons within a Cooper pair can be detected by the electric current cross correlations in the wires. By controlling the relative orientation of the wires, either lithographically or mechanically, on the substrate, the current correlations can be tuned, as dictated by the initial spin entanglement. This proposal of a spin-to-charge readout of quantum correlations is alternative to a recently proposed utilization of the quantum spin Hall insulator.
\end{abstract}


\pacs{72.25.Hg,73.63.Nm,74.78.Na,71.10.Pm}

\maketitle
One of the key features and resources of quantum mechanics is entanglement, particularly in the particle spin sector, which has been an enticing subject since the Einstein-Podolsky-Rosen thought experiment~\cite{einsteinPR35} and, more recently, fueled by the modern proposals for spin-based quantum information processing and computation~\cite{lossPRA98,*divincenzoJMMM99,*zakRNC10}. In order to use an entangled pair of electrons for quantum information technology in a scalable semiconductor setting, it is essential to have a solid-state system that can separate the entangled electrons over appreciable distance. Detecting electron spin entanglement is possible via bunching or antibunching correlations in beam splitters~\cite{burkardPRB00} and transport through Coulomb-blockaded quantum dots forming a Josephson junction~\cite{choiPRB00}. A conceptual headway came with a proposal to spatially separate spin-singlet Cooper pairs (CP's) injected from an $s$-wave superconductor via crossed Andreev reflection (CAR) \cite{deutscherAPL00} in a normal-metal fork \cite{lesovikEPJB01,*recherPRB01}. Later, more elaborate considerations for an $s$-wave superconductor in junction with quantum wires~\cite{benaPRL02,recherPRB02} and quantum dots~\cite{samuelssonPRB04,*sauretPRB04} have been put forward. CAR is essential in all these proposals, and it has been experimentally manifested in the negative nonlocal differential resistance in the system of superconductor in junction with normal metal~\cite{beckmannPRL04,*russoPRL05,*weiNATP10}. CP splitter experiments have been recently performed with quantum dots~\cite{hofstetterNAT09} and carbon nanotubes~\cite{herrmannPRL10}. As another form of CP splitter, we theoretically proposed the system with superconductor straddling a strip of two-dimensional quantum spin Hall insulator (QSHI)~\cite{satoPRL10}, to inject a CP into its gapless edge states. Utilizing the helical Luttinger-liquid (LL) character of the QSHI edges (where each electron moves in the opposite direction to its time-reversed Kramers partner with opposite spin), the spin entanglement can be converted into nonlocal charge-current cross correlations.

In this paper, we consider CP injection into quantum wires with strong spin-orbit interaction (SOI), such as self-doped (and possibly backgated, to control their electron density) InAs nanowires. If only SOI is considered, the spin degeneracy at the $\Gamma$ point ($k=0$) is preserved because of the time-reversal symmetry. However, this degeneracy can be lifted by external magnetic field (facilitated in InAs by a large g factor of electrons). When the chemical potential is set in the corresponding gap at the $\Gamma$ point, gapless states which propagate in the opposite directions with almost opposite spins can be realized at the Fermi points. Note that such a system can closely resemble the helical edge state of the QSHI~\cite{kanePRL05gr,*kanePRL05to,*bernevigPRL06sh}. We consider $s$-wave superconductor connected to a pair of such semiconductor wires in the regime where two CP electrons split into different wires, in the presence of electron-electron repulsion. Effective spin-quantization axes for injecting left- and right-moving electrons into the Fermi points of the two wires are tilted|in one wire relative to the other|by their geometric misalignment. Such tilt affects the current cross correlations in the wires in the way that is similar to  a tunable breaking of the inversion symmetry discussed in Ref.~\cite{satoPRL10}.

At temperatures and voltage bias between the superconductor and the wires that are smaller than the superconductor gap $\Delta$, single-particle injection into the wires is suppressed. In this regime, transport is dominated by the CP tunneling. This process, however, is exponentially suppressed if the distance between the wires exceeds the coherence length of a CP and algebraically on the scale of the Fermi wavelength in the superconductor (depending sensitively on its spatial dimensionality)~\cite{recherPRB02}, posing a potentially serious constraint on the interwire separation. Very importantly, furthermore, if the applied voltage and temperature are smaller than $\Delta$, the parasitic tunneling of two CP electrons into the same wire is suppressed with a power law that is governed by the LL correlations~\cite{recherPRB02}. In this work, we thus focus on the regime where a CP splits ejecting electrons into the different wires. There is a time lag of $\sim\Delta^{-1}$ between such two tunneling events, the longer it is the weaker the LL suppression of the same-wire CP tunneling. However, when two electrons are forced to split and enter different wires at low energies, the leading-order tunneling rates are independent of this time delay (neglecting any interwire interactions)~\cite{recherPRB02}. Therefore, we consider a simplified model with equal-time CP injection of two electrons into two different wires \cite{benaPRL02}. Note also that electron-electron interaction enhances the gap $\Delta$ substantially when the chemical potential is tuned appropriately~\cite{brauneckerPRB10}.

For the wires, Rashba and Dresseslhaus SOI in combination with the Zeeman splitting are considered. Lateral confinement in the wire governs subbands, of which we suppose (at sufficiently low temperature and appropriate backgate bias) only the lowest to be occupied, whose Kramers pairs are split by the lack of both time-reversal and inversion symmetries. In this system, the one-dimensional effective Hamiltonian for a wire oriented along the $x$ axis is given by~\cite{schliemannPRL03,sunPRL07,*gangadharaiahPRB08}
\begin{equation}
\label{H0}
H_{0}=\hbar^2k^{2}/2m^*+\alpha k\hat{\sigma}_{y}+\beta k\hat{\sigma}_{x}-\xi\hat{\sigma}_{z}\,,
\end{equation}
where $m^{*}$ is the effective mass of electron, $\alpha$ ($\beta$) is the strength of the Rashba (Dresselhahus) SOI, and $k$ is the electron wave number. The Dresselhaus part is for the case when a zinc-blende heterostructure is grown in the $[001]$ crystallographic direction, while the wire is oriented in the $[100]$ direction~\cite{scheidPRL08,*wangNT09}. $2\xi={\rm g}\mu_{B}B$ is the Zeeman energy gap at $k=0$, with magnetic field $B$ applied along the $z$ axis, g is the g factor, and $\mu_{B}$ the Bohr magneton. $\hat{\boldsymbol{\sigma}}=(\hat{\sigma}_{x},\hat{\sigma}_{y},\hat{\sigma}_{z})$ are Pauli matrices.

\begin{figure}
\includegraphics[width=0.8\linewidth]{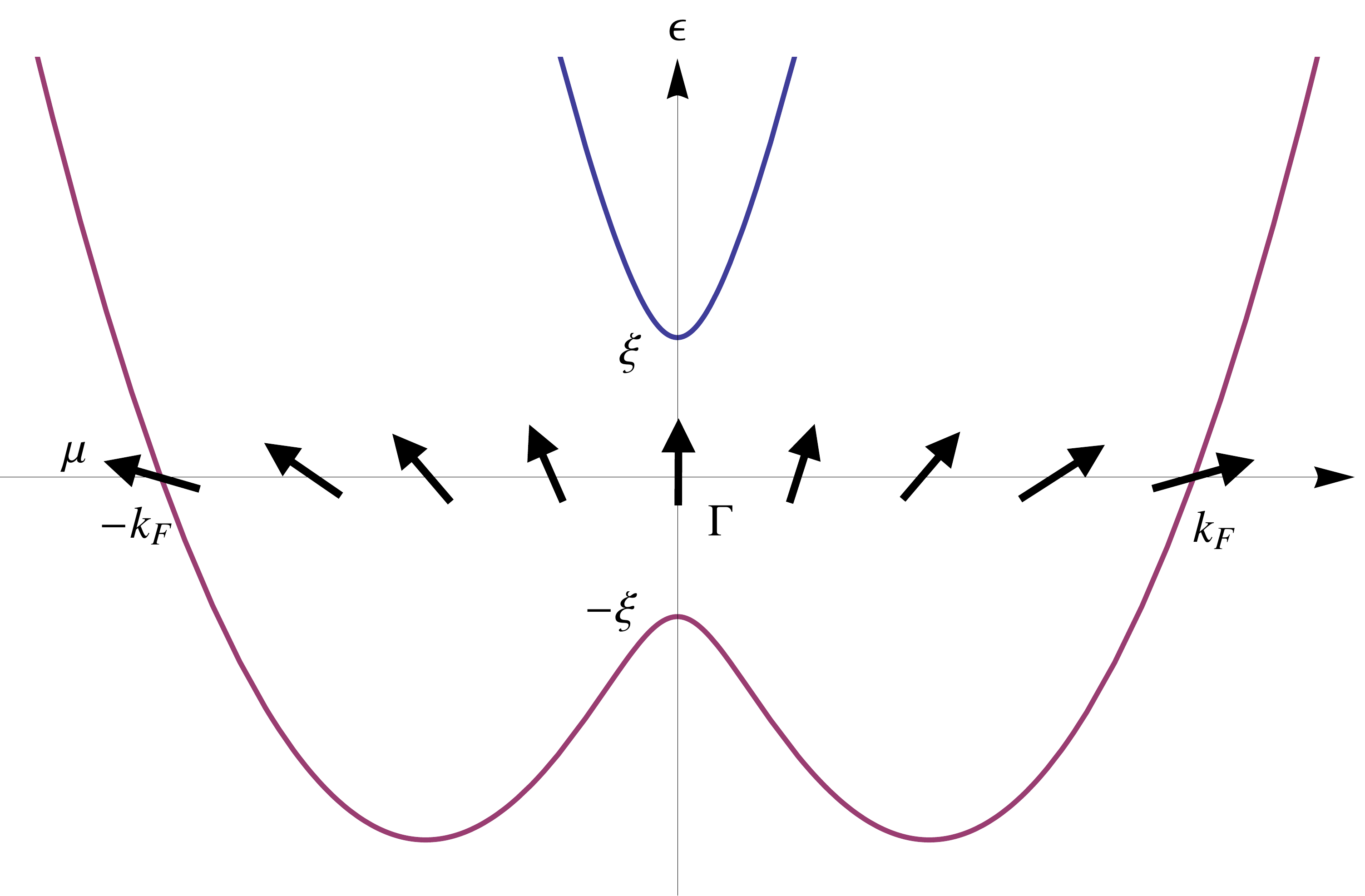}
\caption{Single-particle electron dispersion with Rashba and Dresselhaus SOI. Zeeman splitting $2\xi$ is induced at $k=0$ by a magnetic field in the $z$ direction, and the chemical potential is set in this gap. One-dimensional effective theory is then linearized near $\pm k_{F}$, which define respectively the right- and left-moving electron branches.}
\label{fig1}
\end{figure}

Defining the $k$-dependent effective field $\mathbf{R}(k)=(\beta k,\alpha k,-\xi)$, the Hamiltonian can be written as $H_{0}=\hbar^2k^2/2m+\mathbf{R}(k)\cdot\hat{\boldsymbol{\sigma}}$, and the eigenspinors are found by rotating spinors such that $\mathbf{R}(k)\cdot\hat{\boldsymbol{\sigma}}|\chi_{\pm}(\mathbf{k})\rangle=\pm R(k)|\chi_{\pm}(k)\rangle$, where $R(k)=\sqrt{k^{2}(\alpha^2+\beta^{2})+\xi^{2}}$. The subscripts $+/-$ here label spin up/down along $\mathbf{R}$. The energy eigenstates are thus given by $\psi_{\pm}(k)=\chi_{\pm}(k)e^{i k x}$, with energy $\epsilon_{\pm}(k)=\hbar^{2}k^{2}/2m^{*}\pm R(k)$. The upper and lower $(\epsilon_{+}$ and $\epsilon_{-})$ bands are sketched in Fig.~\ref{fig1}. When the chemical potential $\mu$ is set within the gap, we can linearize the remaining left and right-moving $\epsilon_{-}$ branches within a LL picture. This requires $(eV,k_{B}T)\ll\xi$ and electron-electron interactions that are not strong enough to hybridize the $\epsilon_\pm$ bands. On the other hand, we require the magnetic field to be weak enough on the scale set by $H_c$ of the superconductor (which can be enhanced in mesoscopic structures up to the paramagnetically-limited value of order $\Delta/\mu_B$ \cite{tedrowPRL70}).

\begin{figure}
\includegraphics[width=0.65\linewidth]{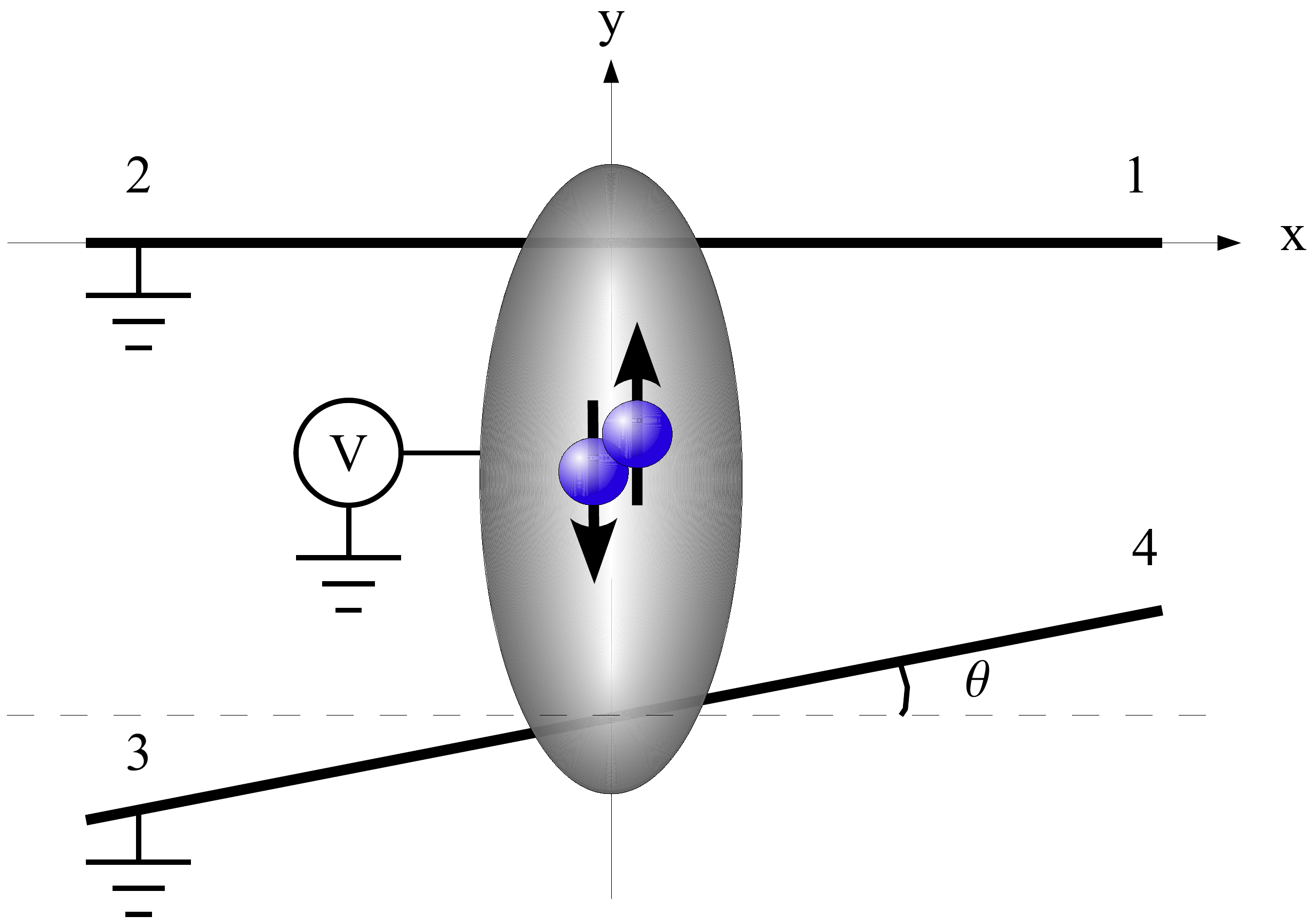}
\caption{$S$-wave superconductor bridging two identical wires. The lower wire is rotated by angle $\theta$ with respect the upper wire. The superconductor is biased by $V$ with respect to the wires.}
\label{fig2}
\end{figure}

Inversion asymmetry between the two wires is introduced by tilting the lower wire (which is otherwise defined along the same crystallographic axis), which rotates the spin quantization axis at each Fermi point of $\epsilon_-$. The upper wire is along the $x$ axis, whereas we suppose the lower wire is placed in the $xy$ plane at an angle $\theta$ with respect to the $x$ axis, as shown in Fig.~\ref{fig2}. (This may in practice be realized by growing both wires parallel to each other on an unstrained crystal, and then distorting the crystal in the $xy$ plane to effectively tilt the wires; depending on the interwire separation, a finite $\theta$ may not require a large strain, whose additional effect on the SOI is neglected.) The SOI in the lower wire is thus given by
\begin{equation}
H_{\rm SO}=\alpha k\left(\cos\theta\hat{\sigma}_{y}-\sin\theta\hat{\sigma}_{x}\right)+\beta k\left(\cos\theta\hat{\sigma}_{x}+\sin\theta\hat{\sigma}_{y}\right)\,.\nonumber
\end{equation}
This changes the effective fields for the upper ($u$) and lower ($d$) wires to
\begin{align}
\mathbf{R}^{(u)}(k)&\equiv\mathbf{R}(k,\theta=0)=(\beta k, \alpha k, -\xi)\,,\nonumber\\
\mathbf{R}^{(d)}(k)&\equiv\mathbf{R}(k,\theta)\nonumber\\
&\hspace{-1cm}=\left[k(-\alpha\sin\theta+\beta\cos\theta),k(\alpha\cos\theta+\beta\sin\theta),-\xi\right]\,.
\label{Rul}
\end{align}
The corresponding Fermi-point eigenspinors and spin splittings are $|\chi^{(u/d)}(\pm k_{F})\rangle\equiv|\chi^{(u/d)}_{r/l}\rangle$ and $\mathbf{R}^{(u/d)}(\pm k_{F})\equiv\mathbf{R}^{(u/d)}_{r/l}$. Note that $\mathbf{R}^{(n)}_{r/l}\cdot\hat{\boldsymbol{\sigma}}|\chi^{(n)}_{r/l}\rangle=- R^{(n)}_{r/l}|\chi^{(n)}_{r/l}\rangle$, and we will assume electronic correlations are not strong enough to significantly affect these Fermi-point spinors. Anticipating tunneling of electrons with well-defined spins from the superconductor into the Fermi points of our wires, we can effectively decompose the fermionic field operators $\psi^{(n)}_\sigma$ ($\sigma=\uparrow/\downarrow$) in terms of the right/left movers $\psi^{(n)}_{r/l}$ in the $n$th wire as~\cite{sunPRL07}
\begin{equation}
\label{psipm}
\psi_{\sigma}^{(n)}=\langle\chi^{(n)}_r|\sigma\rangle\psi^{(n)}_r+\langle\chi^{(n)}_l|\sigma\rangle\psi^{(n)}_l\,.
\end{equation}
The full wire Hamiltonian (\ref{H0}) is bosonized \cite{giamarchiBOOK04} near the Fermi points to give an essentially helical (so long as the Zeeman term $\xi$ is weak) LL  \cite{stromPRL09}:
\begin{equation}
H_{0}=v\sum_{n=u,d}\int{\frac{dx}{2\pi}\left[\frac{1}{g}\left(\partial_{x}\phi^{(n)}\right)^{2}
+g\left(\partial_{x}\theta^{(n)}\right)^{2}\right]}\,,\nonumber
\end{equation}
where $\phi^{(n)},\theta^{(n)}=(\phi^{(n)}_r\pm\phi^{(n)}_l)/2$ obey commutation relations $[\theta^{(n)}(x),\phi^{(m)}(y)]=(i\pi/2){\rm sgn}(x-y)\delta_{nm}$. $\phi_{r/l}^{(n)}$ parametrize fermionic operators as $\psi_{r/l}^{(n)}\propto e^{\pm i\phi^{(n)}_{r/l}}$.

The tunneling Hamiltonian, which describes nonlocal injection of the spin singlet CP from an $s$-wave superconductor into the two quantum wires is given by \cite{benaPRL02}
\begin{equation}
\label{tunnel}
H_{T}=\Gamma e^{-2eVt}\left[\psi^{(u)}_{\uparrow}\psi^{(d)}_{\downarrow}(0)-\psi^{(u)}_{\downarrow}\psi^{(d)}_{\uparrow}(0)\right]+\rm{H.c.}\,.
\end{equation}
In this model, two electrons from a singlet CP split and tunnel simultaneously into the upper and lower wires at their respective origins. $V$ is the voltage applied between the superconductor and the wires, which is set to be smaller than $\Delta$ to preclude quasiparticle excitations. Expanding spin-dependent operators $\psi^{(n)}_{\uparrow/\downarrow}$ in terms of the chiral modes pertinent to the wires, Eq.~(\ref{psipm}), we can rewire tunneling Hamiltonian (\ref{tunnel}) according to
\begin{equation}
\psi^{(u)}_{\uparrow}\psi^{(d)}_{\downarrow}-\psi^{(u)}_{\downarrow}\psi^{(d)}_{\uparrow}
=\sum_{\mu,\nu=r,l}K_{\mu\nu}\psi^{(u)}_{\mu}\psi^{(d)}_{\nu}\,,\nonumber
\end{equation}
where $K_{\mu\nu}$ are the complex-valued expansion coefficients.

Finally, current-current correlations at the four end points of the two wires in Fig.~\ref{fig2} are considered. The symmetrized noise spectrum,
\begin{equation}
\label{S}
S_{ij}(\omega)=S_{ji}(-\omega)=\int_{-\infty}^{\infty}dte^{i\omega t}\langle\{\delta I_{i}(t),\delta I_{j}(0)\}\rangle\,,
\end{equation}
is calculated using Keldysh formalism \cite{chamonPRB96,satoPRL10}. Here $\delta I_{i}(t)=I_{i}(t)-\langle I_{i}(t)\rangle$ are the current fluctuations, $i$ labeling four outgoing channels in the wires ($i=1$, upper right; 2 upper left; 3, lower left; and 4, lower right branches). See Fig.~\ref{fig2}. These currents are given in the bosonic representation by $I^{(n)}=(vg/\pi)\partial_{x}\theta^{(n)}$. Using Eqs.~(\ref{psipm})-(\ref{S}), current correlations are calculated in terms of $K_{\mu\nu}$, with the final answer (reflecting spin-rotational symmetry of a singlet CP) depending only on $|K_{\mu\nu}|^{2}$,
\begin{equation}
\label{Ksq}
|K_{\mu\nu}|^{2}=\left(1-\hat{\mathbf{R}}^{(u)}_{\mu}\cdot\hat{\mathbf{R}}^{(d)}_{\nu}\right)/2\,.
\end{equation}
Here, $\hat{\mathbf{R}}_{\mu}^{(n)}=\mathbf{R}_{\mu}^{(n)}/R_{\mu}^{(n)}$, and $|K_{\mu\nu}|^2$ can be evaluated using Eq.~(\ref{Rul}). $R^{(n)}_{\mu}=\sqrt{k_{F}^{2}(\alpha^{2}+\beta^{2})+\xi^{2}}$, for $n=u,l$ and $\mu=\pm$, independent of the orientation of the wire or electron chirality. Furthermore, since $\mathbf{R}^{(u)}_{\mu}\cdot\mathbf{R}^{(d)}_{\nu}=\mu\nu k_{F}^{2}(\alpha^{2}+\beta^{2})\cos\theta+\xi^2$ (identifying $r=+$ and $l=-$), we find that $|K_{++}|^2=|K_{--}|^2$ and $|K_{+-}|^2=|K_{-+}|^2$.
Lumping Zeeman and SOI energies into a dimensionless parameter $\lambda=\xi/k_{F}\sqrt{\alpha^{2}+\beta^{2}}$, we finally arrive at a rather simple expression for Eq.~(\ref{Ksq}):
\begin{equation}
\label{KLO}
|K_{\mu\nu}|^{2}=(1+\lambda^{2})^{-1}(1-\mu\nu\cos\theta)/2\,,
\end{equation}
Using these $K_{\mu\nu}$, the following expressions for the noise spectra are obtained at zero frequency ($\omega=0$):
\begin{align}
\label{noise}
S_{13}&=S_{31}=S_{24}=S_{42}=eI(1+g^{2}\cos\theta)/4\equiv S_+\,,\nonumber\\
S_{14}&=S_{41}=S_{23}=S_{32}=eI(1-g^{2}\cos\theta)/4\equiv S_-\,,\nonumber\\
S_{11}&=S_{22}=S_{33}=S_{44}=eI(1+g^{2})/4\,,\nonumber\\
S_{12}&=S_{21}=S_{34}=S_{43}=eI(1-g^{2})/4\,,
\end{align}
where
$I=G(eV/\epsilon_{F})^{2\gamma}V/(1+\lambda^2)$, with $\gamma=(g+g^{-1})/2-1$, is the total tunneling current from the superconductor to the wires (at $k_BT\ll|eV|$; $G\propto|\Gamma|^2$ is proportional to the CAR conductance in the absence of LL correlations) \cite{benaPRL02,satoPRL10}. This current vanishes in the limit $\lambda\gg1$, when both wires become fully spin polarized thus blocking the CP tunneling. Notice that the magnetic field did not scramble helical structure of the interwire cross correlations, which turn out to be the same [apart from the overall suppression by $(1+\lambda^2)$] as in the case of the time-reversal symmetric QSHI \cite{satoPRL10}. This is one of the key results of this paper.

The interwire cross-correlation spectra (\ref{noise}) are given by
\begin{equation}
\label{noisepm}
S_{\pm}(\theta,\lambda)\propto(1+\lambda^2)^{-1}(1\pm g^{2}\cos\theta)\,,
\end{equation}
which are modified from those in Ref.~\cite{satoPRL10} only by the magnetic-field suppression factor of $(1+\lambda^2)^{-1}$. In Ref.~\cite{satoPRL10}, the angle $\theta$ dependence for the CP injection into the helical edge states of a QSHI is due to a tunable asymmetry between two edges (induced by a local application of strain or gate voltage to an otherwise inversion-symmetric system). Here, $\theta$ dependence comes from the mechanical rotation of the lower wire by the angle $\theta$. Notice that the definitions for $S_{+}$ and $S_{-}$ are interchanged here in comparison to Ref.~\cite{satoPRL10}. This is because the quantum wires considered here do not have the inversion symmetry of helical edge states on the opposite sides of a QSHI strip. Despite this fundamental difference, we can clearly see the same structure in the CP noise cross correlations for both the present quantum-wire system and the helical QSHI edges. According to Eq.~(\ref{noisepm}), we can extract the LL interaction parameter $g$ (which is typically $g\sim0.1-1$ \cite{auslaenderSCI02,*auslaenderSCI05} in semiconducting wires) from the interwire cross correlations: $g^2\cos\theta=(S_{+}-S_{-})/(S_{+}+S_{-})$. While in Ref.~\cite{satoPRL10} the angle $\theta$ is a parameter that may not be precisely known, in the present set-up the rotation angle $\theta$ of the lower wire can be experimentally well defined, so that $g$ can be found by measuring $S_{+}$ and $S_{-}$ for an arbitrary value of $\theta$ that is away from $\pi/2$.

In the discussion so far, we were considering only one specific crystallographic orientation of the wires. Namely, the heterostructure growth is in the $[001]$ crystallographic direction and each wire is defined (e.g., electrostatically) along the $[100]$ direction. However, while the Rashba SOI is rotationally invariant around the normal axis, the Dresselhaus SOI is sensitive to the wire orientation on a crystal's surface \cite{meierNATP07,scheidPRL08}. Suppose that with the same crystal growth direction of $[001]$, the wire is defined at an angle $\theta_{D}$ from the $[100]$ direction. In this case, the Dresselhaus SOI part of the Hamiltonian is given by \cite{scheidPRL08}
\begin{equation}
H_{D}=\beta k\left[\cos(2\theta_{D})\hat{\sigma}_{x}-\sin(2\theta_{D})\hat{\sigma}_{y}\right]\,.\nonumber
\end{equation}
This crystallographic orientation and the associated Hamiltonian are now chosen for the upper wire, with our coordinate system still placed (as in Fig.~\ref{fig2}) with the $x$ direction collinear with the wire. The corresponding effective-field vector is then $\mathbf{R}^{(u)}(k)=[\beta k\cos(2\theta_{D}),\alpha k-\beta k\sin(2\theta_{D}),-\xi]$. Since the lower wire is rotated in the $xy$ plane by the angle $\theta$ with respect to the upper wire, $\mathbf{R}^{(l)}$ obtained by the corresponding rotation on $\mathbf{R}^{(u)}$ is given by $\mathbf{R}^{(l)}=[-\alpha k\sin\theta+\beta k\cos(2\theta_{D}-\theta),\alpha k\cos\theta-\beta k\sin(2\theta_D-\theta),-\xi]$. The absolute value of $\mathbf{R}^{(u/l)}$ is modified by $\theta_{D}$: $R^{(u/l)}=\sqrt{k^2[\alpha^2+\beta^2-2\alpha\beta\sin(2\theta_{D})]+\xi^2}$. Both the direction and the magnitude of $\mathbf{R}^{(u/l)}$ are thus modified, affecting $K_{\mu\nu}$ in Eq.~(\ref{Ksq}). We still have $|K_{++}|^{2}=|K_{--}|^{2}$ and $|K_{+-}|^{2}=|K_{-+}|^{2}$ according to Eq.~(\ref{Ksq}). In fact, the modification of $|K_{\mu\nu}|^2$ can be absorbed by redefining $\lambda$ entering Eq.~(\ref{KLO}) as $\lambda=\xi/k_{F}\sqrt{\alpha^2+\beta^2-2\alpha\beta\sin(2\theta_{D})}$, with all subsequent relations for the noise spectra unmodified. In particular, apart from the modified geometric spin factor $\lambda$, which suppresses the overall strength of the CAR, $S_{\pm}$ in Eq.~(\ref{noisepm}) remain the same. This means we can choose any wire orientation on the crystal surface without altering the essence of the noise cross correlations. One special point is $\theta_D=\pi/4$ when $\alpha=\beta$ (or $\theta_D=-\pi/4$ when $\alpha=-\beta$), corresponding to the ``persistent spin helix" \cite{schliemannPRL03,bernevigPRL06su}, where $\lambda$ blows up and the CAR is fully blocked (reflecting exact cancellation of the SOI terms).

Let us also comment on a possible triplet pairing of the injected electrons, e.g., if the two terms in the tunneling Hamiltonian \eqref{tunnel} acquire a relative phase difference: $e^{i\delta/2}\psi^{(u)}_{\uparrow}\psi^{(d)}_{\downarrow}-e^{-i\delta/2}\psi^{(u)}_{\downarrow}\psi^{(d)}_{\uparrow}$. We can rewrite it as $\cos(\delta/2)(\psi^{(u)}_{\uparrow}\psi^{(d)}_{\downarrow}+\psi^{(u)}_{\downarrow}\psi^{(d)}_{\uparrow})-i\sin(\delta/2)(\psi^{(u)}_{\uparrow}\psi^{(d)}_{\downarrow}+\psi^{(u)}_{\downarrow}\psi^{(d)}_{\uparrow})$. The corresponding modification of $|K_{\mu\nu}|^{2}$ in Eq.~(\ref{KLO}) can be accounted for by the replacement $\theta\to\theta-\delta$, with the same $\delta$ shift of $\theta$ appearing in the noise expressions. Interestingly, the phase difference in the tunneling terms has the same effect on the current correlations as a mechanical rotation of the wires. Such triplet component in tunneling can be effectively induced by tunneling away from the Fermi points at finite temperature and/or voltage, and artificially enhanced in more complex tunneling setups \cite{virtanenCM11}.

Another concern to be mentioned is that, if the superconductor is in a slab shape, the perpendicular critical field is reduced. This issue can be mitigated by applying in-plane magnetic field. For the case of a thin-film superconductor, the critical field is further enhanced (up to its paramagnetic limit \cite{tedrowPRL70}) when the magnetic penetration depth is greater than its thickness. $\mathbf{R}$ in Eq.~(\ref{Rul}) needs to be modified accordingly. Since the magnetic-field and SOI contributions to $\mathbf{R}$ are not perpendicular to each other any more, the resulting energy band is not symmetric as in Fig.~\ref{fig1}. In turn, $|K_{\mu\nu}|^{2}$ in Eq.~(\ref{KLO}) and the formula in Eq.~(\ref{noise}) acquire some corrections. In the limit of $\lambda\ll1$, the corrections are small, however, and we recover the same noise behavior as in Eq.~(\ref{noisepm}). In the strong-field limit, $\lambda\gtrsim1$, on the other hand, a more careful analysis would be warranted.

Now let us return to Eq.~(\ref{noisepm}) to see the feasability of this theory in an experiment. A very large magnetic splitting (on the scale of the SOI) in the wires, $\lambda\gg1$, blocks Andreev reflection \cite{jongPRL95}, when the Fermi level is inside the $\Gamma$-point gap. The SOI is large in the InAs-based heterostructures and wires, where the Rashba parameter is $\alpha\lesssim10^{-11}$~eV\,m (being tunable by electrostatic gating) \cite{nittaPRL97,*kogaPRL02,*dharaPRB09,*estevezPRB10}, $\beta\ll\alpha$, and g factor is $\approx15$. For electron densities in the range of $10-100$~$\mu{\rm m}^{-1}$, this gives for the magnetic field $B\sim0.1-1$~T corresponding to $\lambda\sim1$. Both $\alpha$ and g factor can be considerably lower (both up to two orders of magnitude) in InGaAs-based heterostructures, which can make also $\alpha\sim\beta$ \cite{meierNATP07}, while the corresponding magnetic field range remains roughly the same. This gives us a favorable operational bound for the magnetic field, which opens the $\Gamma$-point gap without compromising the strength of the CAR, while also not exceeding the paramagnetically-limited critical field (with $T_c\gtrsim1$~K). Taking everything into account, this means the experiments can be done at temperatures close to 1~K.

This work was supported by the Alfred P. Sloan Foundation, the NSF under Grant No. DMR-0840965 (Y.T.), and by the Swiss NSF and DARPA QuEST (D. L.).

\end{document}